\documentclass[12pt]{elsart}
\usepackage{graphicx}

\begin{document}
\bibliographystyle{elsart-num}
\begin{frontmatter}
\title{The MINOS Light Injection Calibration System}

\author[sussex]{P. Adamson\thanksref{ucl2}}
\author[ral]{J. Alner}
\author[ucl]{B. Anderson}
\author[umn]{T. Chase}
\author[ucl]{P.J. Dervan\thanksref{sheffield}}
\author[ral]{T. Durkin}
\author[sussex]{E. Falk}
\author[sussex]{P.G. Harris\corauthref{me}} \ead{P.G.Harris@sussex.ac.uk}
\corauth[me]{Corresponding author.}
\author[caltech]{D.G. Michael}
\author[sussex]{R. Morse}
\author[ucl]{R. Nichol}
\author[ucl]{R. Saakyan}
\author[ucl]{C. Smith}
\author[sussex]{P.N. Smith}
\author[ucl]{J. Thomas}
\author[utex]{R. Webb}
\author[sussex]{R.F. White}

\address[sussex]{Department of Physics, University of Sussex, 
Falmer, Brighton BN1 9QJ, UK} 
\address[ral]{Rutherford Appleton Laboratory, Chilton, Didcot
OX11 0QX, UK}
\address[ucl]{Department of Physics, University College London, 
Gower Street, London WC1E 6BT, UK}
\address[umn]{Mechanical Engineering Department, University of 
Minnesota, 111 Church St SE, Minneapolis MN  55455, USA}
\address[caltech]{California Institute of Technology High 
Energy Physics, Charles C. Lauritsen Laboratory,
Pasadena CA 91125, USA }
\address[utex]{Texas A \& M University Physics Department, 
College Station, TX 77843-4242 }

\thanks[ucl2]{Present address: Department of Physics, University 
College London, Gower Street, London WC1E 6BT, UK}
\thanks[sheffield]{Present address: Department of Physics \& 
Astronomy, Hicks Building, University of Sheffield, Hounsfield 
Road, Sheffield S3 7RH, UK}



\begin{abstract}
A description is given of the light-injection calibration system 
that has been developed for the MINOS long-baseline neutrino 
oscillation experiment.  The system is based upon 
pulsed blue LEDs monitored by PIN photodiodes.  It is designed to 
measure non-linearities in the PMT gain curves, as well as 
monitoring any drifts in PMT gain, at the 1\% level. 

\end{abstract}

\end{frontmatter}

\section{MINOS overview}
The MINOS experiment \cite{tdr} will investigate the phenomenon of 
neutrino oscillations \cite{pontecorvo58,pontecorvo67,bilenky87}  
by sending a beam of muon neutrinos 735 km through the Earth from 
the Fermi National Accelerator Laboratory (FNAL), near Chicago, to 
a detector in the Soudan Underground Mine State Park in northern 
Minnesota.  The experiment uses three detectors:
\begin{enumerate}
\item{} The Near Detector (ND) at FNAL.  This will determine the 
characteristics of the neutrino beam before it begins its journey. 
Only neutrino events and cosmic ray muons are measured here; there 
is no ``test beam'' available.  
\item{} The Far Detector (FD), situated 713 m underground in the 
Soudan mine, where the neutrino beam emerges from the Earth.  Like 
the near detector, this measures only neutrino events and cosmic 
ray muons.
\item{} The Calibration Detector (CD) at CERN is a small detector, 
similar in structure to both near and far detectors, that is being 
exposed to particle beams at CERN to study its response to 
electrons, muons and hadrons of known energies.  
\end{enumerate}

\section{MINOS detectors}

The MINOS far detector is an 8 m wide octagonal tracking 
calorimeter, consisting of 486 layers of one-inch-thick steel 
interleaved with scintillator, giving a total mass of 5.4 
kilotons. It has a toroidal magnetic field of strength 
approximately 1.3 T.

Each plane of scintillator consists of 192 strips, each 4 cm 
wide and 1 cm thick with a co-extruded TiO$_{2}$ cladding to 
maximise internal reflection.  Light from these strips is 
wavelength-shifted from blue to green and carried within 
acrylic fibres \cite{kuraray_wls} to Hamamatsu \cite{hamamatsu}
multi-anode photomultiplier tubes situated around the edge of the 
detector.  Scintillator strips are in orthogonal directions in 
alternate planes. This gives a 3D readout with an energy 
resolution $\sigma/E$ of approximately $23\%/\sqrt{E}$ 
for electromagnetic showers and $55\%/\sqrt{E}$ for hadronic 
showers.  The magnetic  field allows measurement of the muon 
momentum by curvature (14\% resolution for a 10 GeV muon) as 
well as by range for stopping muons (giving approximately 6\% 
resolution). 

\subsection{Calorimeter energy calibration}

In order to measure the neutrino mixing parameters 
$\Delta m^2$ and $\sin^2(2\theta)$ precisely, the 
energy response of the detectors has to be calibrated to 
within a few percent. The goal is a 5\% absolute measurement of 
the energy in each detector, and consistency to within 2\% 
between the near and far detectors.  This is carried out in 
several stages \cite{micawg}:
\begin{itemize}
        \item{} A light-injection calibration system is required 
to monitor, and to compensate for, changes in amplification 
(and non-linearity) of the photomultiplier tubes used in the 
detectors.  The light-injection system is the principal subject 
of this paper.
        \item{} Cosmic-ray muons are used to perform the 
strip-to-strip calibration within each detector, and the relative 
calibration between the near and far 
detectors \cite{litchfield,numi694}.  
The strip-to-strip calibrations do not require a high 
statistical precision ($\sim10\%$ per strip per month), as both 
muon tracks and hadronic showers contain many strips.  There are 
expected to be about 530 cosmic ray hits/strip/month in the far 
detector, giving an accuracy of 2.4\% per strip per month, and 
about 25,000 hits/strip/month in the near detector, giving an 
accuracy there of 0.3\%/strip/month.
        
        Stopping muons can be used to provide the relative 
calibration of  near to far detectors.  There are estimated to 
be 700 stopping muons per day in the far detector, and two 
orders of magnitude more than this in the near detector.  This 
yields a 1\% statistical precision for the near-far detector 
calibration each day. 
        \item{} A study to compare the energy deposition by 
muons with that from hadrons and hadronic showers is being 
carried out with the calibration detector in a charged-particle 
test beam at CERN.  
\end{itemize}

\subsection{Optical readout}

\label{sec:det}

\begin{figure}[htbp]
  \begin{center}
    \resizebox{0.8\textwidth}{!}{\includegraphics{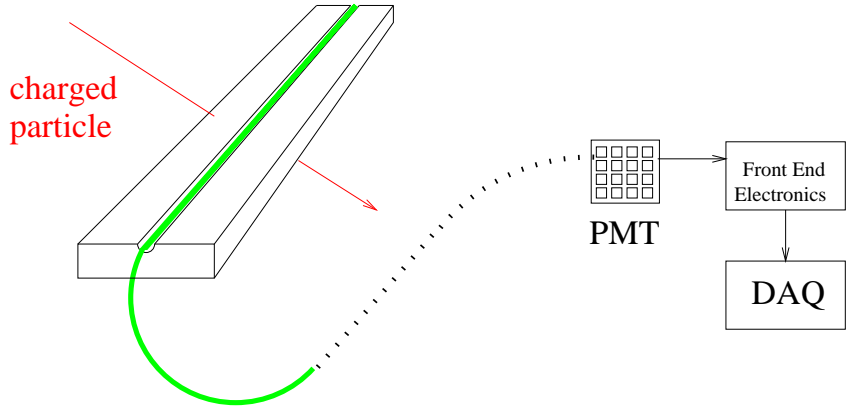}}
    \caption{Signal readout path from scintillator to electronics 
in MINOS.  The green acrylic wavelength-shifting fibre lies 
within a groove running down the centre of the scintillator 
strip.  Light is carried within this fibre to the edge of the 
detector, and then via clear acrylic fibre to the 
photomultiplier tube.}
    \label{fig:readout}
  \end{center}
\end{figure}

The MINOS readout chain is illustrated schematically in 
Fig.~\ref{fig:readout}. 
A 1.2 mm Kuraray Y11 wavelength-shifting fibre 
\cite{kuraray_wls} lies in a groove along the length of each 
scintillator strip. Between 16 and 28 such strips are enclosed 
in a light-tight aluminium case, known as a module.  There are 
eight modules in a full plane.
 
The wavelength-shifting fibres run to a connector at the end of 
the aluminium case.  Clear fibres \cite{kuraray_clear}
then carry the light to boxes containing Hamamatsu multi-anode 
photomultiplier tubes.

\begin{figure}[htb]
  \begin{center}
    \resizebox{0.6\textwidth}{!}{\includegraphics{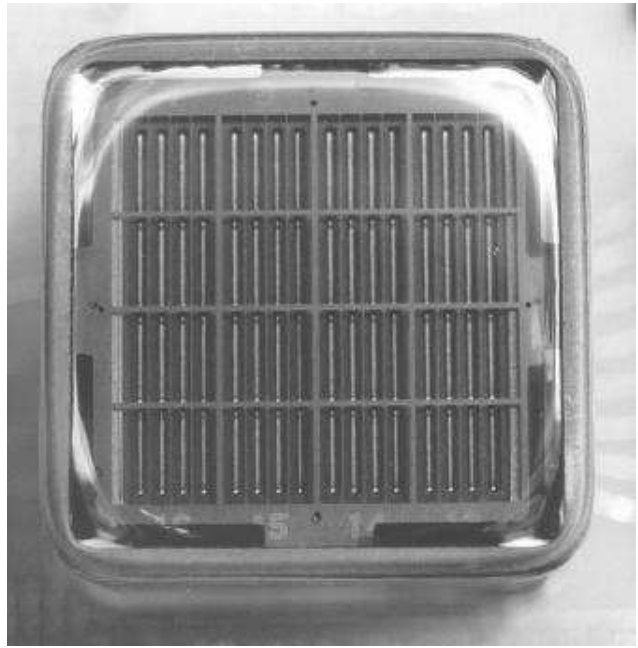}}
    \caption{The face of a Hamamatsu R5600-M16 multi-anode 
photomultiplier as used in the far detector. Photograph 
courtesy K. Lang.}
    \label{fig:m16}
  \end{center}
\end{figure}

For the far detector, Hamamatsu M16 photomultipliers 
\cite{lang01} are used  (see Fig.~\ref{fig:m16}), with
fibres from eight separate scintillator strips feeding each 
PMT pixel. This ``optical multiplexing'', 
which provides a substantial saving in the cost of readout 
electronics, is possible because the event rate at the far 
detector is so low that there will never be more than one event 
in the detector at any one time. The typical width of a shower 
event in MINOS is 50 cm, so the 1 m spacing between strips 
illuminating a single pixel is appropriate. The multiplexing 
pattern is slightly different on each side of the detector
in order to aid deconvolution.

The far-detector readout electronics uses a modified version of 
the Viking VA chip \cite{toker94} made by
IDE AS of Norway \cite{ideas}. This is an integrated circuit 
containing 32 channels of shaping amplifiers and track-and-hold 
circuitry.  When triggered, each channel is digitised in turn. 
The data are then transferred to VME
readout control crates, with single-board computers running the
VxWorks real-time operating system.

Instantaneous event rates in the near detector will be 
$5 \times 10^5$ higher than in the far detector.  The near 
detector  therefore uses QIE \cite{zimmerman96} 
electronics, a high-speed, multi-ranging ADC system with no 
dead time.  Its principal difference from the far
detector electronics is that it divides input signals into
19 ns time buckets, and digitises each separately. These
separate digitisations are recombined during the event-building
process.  The light-injection calibration system incorporates 
an external trigger that can be used to produce optical pulses 
at different starting positions within the 19 ns
time bucket, in order to investigate any possible systematic 
differences in response at the bucket boundaries.  

The near detector is not optically multiplexed (although its 
muon spectrometer section is electronically multiplexed).  It 
uses Hamamatsu M64 (64-anode) photomultiplier tubes, and 
has a total of 12,000 readout channels. Unlike the far 
detector, only one end of each strip is read out: the other 
end is terminated with a reflector to maximise the detected light.  The calibration detector will use both near and far detector readout systems, in order to carry out a systematic comparison of the responses of each.

\section{Light-injection system overview}
The light-injection system is based upon pulsed blue 
light-emitting diodes (LEDs).  A rack-mounted box, known as a 
``pulser box'', contains a set of 20 LEDs, each of which has 
an optical fan-out allowing it to illuminate multiple 
individual fibres.  These fibres carry light to a set of 
optical connectors on the back panel of the pulser box.

From the pulser box, optical fibres carry the light to the 
outer edges of the MINOS detector.   A set of highly-reflective 
cavities situated there -- the ``light-injection modules'' -- 
allows the blue LED light to illuminate the green 
wavelength-shifting fibres, thus producing pulses of light 
that mimic the signals from the scintillator.  

The intensity of injected light is monitored by PIN 
photodiodes that are read out simultaneously with the PMTs.

In the following sections, the requirements of and 
constraints upon the light-injection system are discussed, 
after which each component of the system is examined in detail.

\section{Light-injection system requirements}
\subsection{Aims}
\label{sec:aims}
The purpose of the light injection system is fourfold:
\begin{enumerate}
\item Mapping the gain curve.  The response of the 
photomultiplier tubes becomes non-linear at light levels 
exceeding about 100 photoelectrons \cite{lang01}.  The light 
injection system is designed to be able to characterise the 
response of every optical readout channel on a regular basis 
(up to once per month in routine operation).
\item Monitoring short-term gain drifts.  The gain of a 
photomultiplier tube is sensitive to small changes in 
temperature and in applied high voltage.  By injecting light 
at a single point on the gain curve on an hourly basis, any 
resulting changes in gain can be monitored closely and 
compensated for.
\item Evaluating PMT gain.  Pulsing at very low light levels 
allows the acquisition of single photoelectron spectra for each 
channel, from which the gain can easily be calculated.  
\item Confirming optical path integrity.  The light injection 
system provides a quick and thorough check to ensure that no 
fibres are broken.
\end{enumerate}
\subsection{Design constraints}
A number of constraints were imposed upon the design of the 
light injection system:
\begin{enumerate}
\item The amount of light applied to any pixels on a given PMT 
that are illuminated simultaneously should differ by no more 
than a factor of two.  This helps to avoid excessive 
non-linearities and possible problems from crosstalk when 
injecting light towards the top end of the dynamic range.  
It is also useful in reducing the amount of data required in 
order to cover the dynamic range for all pixels, as one does 
not then have dimly and brightly illuminated pixels that require 
separate calibration operations at different levels of LED 
brightness.
\item The M16s used on the far detector are multiplexed, 
with eight fibres carrying light to each pixel. No two fibres 
on any pixel may be illuminated simultaneously. Adjacent pixels 
should also be avoided if possible, to minimise crosstalk 
problems.
\item The dynamic range of light injected must range from a 
fraction of a photoelectron up to several hundred 
photoelectrons on all pixels. Furthermore, this dynamic range 
should be covered with a step size of the order of one 
photoelectron.
\item The relative levels of light injected must be measured 
by a separate device having a response that is linear to 1\%, 
in order to characterise the non-linear response of the PMTs.  
PIN photodiodes were selected for this purpose.  
\item The response of the PIN photodiodes must be stable 
to within 1\% over the timescale of a month.
\item The relative arrival times of the optical pulses at 
each PMT must be stable to within approximately a nanosecond.
\end {enumerate}

\section{Light-emitting diodes}
\subsection{LEDs vs.\ lasers}
Previous experiments needing light-injection calibration of the 
type required for MINOS have often used scintillator excited 
by pulsed lasers, with a variety of neutral-density filters to 
provide the required dynamic range in light intensity.   The 
development over recent years of the so-called ``ultra-bright'' 
blue LEDs enabled a simpler, solid-state system to be considered.  
Lasers retain three advantages: they are several orders of 
magnitude brighter than even the brightest LEDs; the pulses 
are extremely fast, which can better mimic the pulses produced 
by the passage of charged particles through scintillator; and 
the spectrum of light produced by the scintillator is almost 
identical for data and for calibration pulses.  For MINOS, 
however, these possible benefits of lasers were outweighed 
by the many other advantages of LEDs.  LEDs exhibit minimal 
pulse-to-pulse fluctuations in intensity, so the response of 
the PMT is dominated entirely by photoelectron statistics.  In 
an LED-based system, there are no moving parts to wear out or 
break, so the system is appropriately robust for long-term 
remote operation. In addition, there are no laser parts requiring 
regular replacement,  servicing or gas supplies; the light 
intensity can be changed by software, instead of relying upon 
neutral-density filters which may become burned (and therefore 
non-linear) and which also can only be adjusted slowly; and 
high repetition rates are possible, so calibration can be 
carried out quickly.  The temporal profile of the optical 
pulse can be changed to some extent, to explore any effects 
thereof.   Finally, the LED system costs less than a comparable 
laser-based system.  A major factor driving the cost difference 
is within the fibre plant: blue LED light can propagate down 
polymer fibre which has a very high absorption at the UV 
wavelengths generally produced by lasers.
         
The LEDs used in the pulser system are wide-angle Hewlett-Packard
HLMP-CB30 \cite{hp}. Initial development 
work had been done with the LEDs made by Nichia Chemical 
Industries \cite{nichia};  the change to the HP LED was forced 
as Nichia altered the specification of their LED
to one having a much slower response, which made the pulses 
unacceptably long. 

\subsection{Maximum current}
A nominal maximum current specified by the manufacturer of 
100 mA is determined by heating effects, which can damage 
the plastic lens in which the LED is set; however, this is 
not a problem with the extremely low duty cycle in the MINOS 
pulser circuit (some tens of nanoseconds per millisecond), 
and tests showed no sign of damage with currents as high 
as 500 mA. The MINOS pulser circuit uses a maximum current 
of 200 mA; above this, the visible light output begins to 
saturate, and the quality of the pulse shape deteriorates.  

\subsection{Lifetime}

None of the LEDs tested have ever reached the limit of 
their lifetime.  In one instance, the tests included pulses 
equivalent to 2,000 years of normal MINOS operation, with 
no adverse effects observed.

\subsection{Comparison of spectra}

\begin{figure}[thb]
  \begin{center}
    \resizebox{0.8\textwidth}{!}{\includegraphics{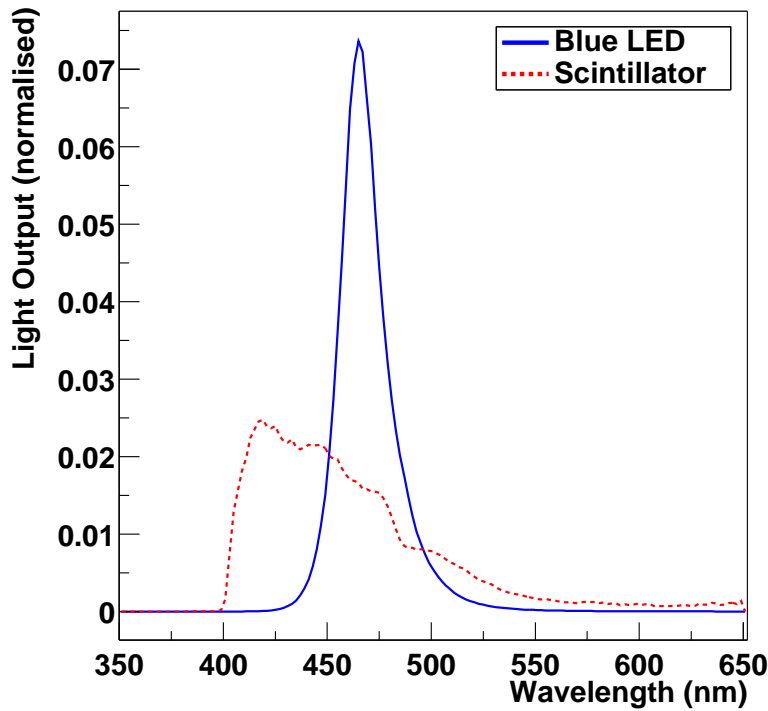}}
    \caption{A comparison of the wavelengths of light emitted 
by the blue LED and stimulated MINOS scintillator.}
    \label{fig:lscomp}
  \end{center}
\end{figure}

\begin{figure}[hbt]
  \begin{center}
    \resizebox{0.8\textwidth}{!}{\includegraphics{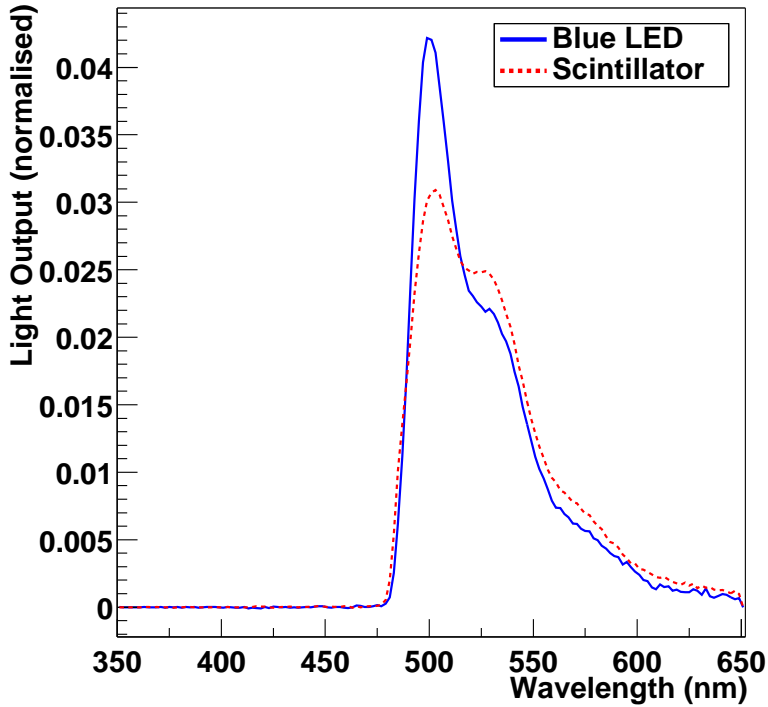}}
    \caption{Spectrum of light emitted from wavelength-shifting 
fibre when excited with the light in Fig.~\ref{fig:lscomp}.}
    \label{fig:lswlscomp}
  \end{center}
\end{figure}

Fig.~\ref{fig:lscomp} shows a comparison of the spectra of light
emitted from the blue LED with light from an excited piece of 
MINOS scintillator \cite{thomas99}. The spectra are quite 
different. When the light is absorbed and re-emitted by the 
wavelength-shifting fibre, however, the spectra are far better 
matched, as shown in Fig.~\ref{fig:lswlscomp}.  The measurement 
was made using a 300 mm piece of WLS fibre, half of which was 
glued into a groove in a 150 mm length of MINOS scintillator 
in the usual manner.  The laser light was injected approximately 
halfway along the scintillator; the LED illuminated the WLS 
fibre at a point adjacent to the end of the scintillator 
(and thus approximately halfway along the WLS fibre).   Light 
from the end of the WLS fibre was transported through a 
quartz fibre to a monochromator and calibrated photomultiplier 
tube.  In MINOS, the situation will be slightly different: 
most signal pulses will have a much longer path through the green 
fibre, where self-absorption reduces the short-wavelength 
component of the light \cite{border01}, whereas the blue 
LED light will be absorbed at the very end of the green fibre.  
Nonetheless, the similarity of the spectra eliminates any 
concerns about wavelength-dependent ageing of the
photocathode (which, in any case, should occur on a timescale 
much longer than the one-month period over which the 
light-injection system is required to monitor gain stability).

\subsection{Pulse shape}
A short, intense pulse of light in a PMT does not necessarily 
yield the same integrated charge as an identical amount of 
light delivered over a longer time, and so it is important that 
the shape of the optical pulse arriving at the photocathode 
from light injection should be  similar to that from the 
detection of charged particles.  

\begin{figure}[hbt]
  \begin{center}
    \resizebox{\textwidth}{!}{\includegraphics{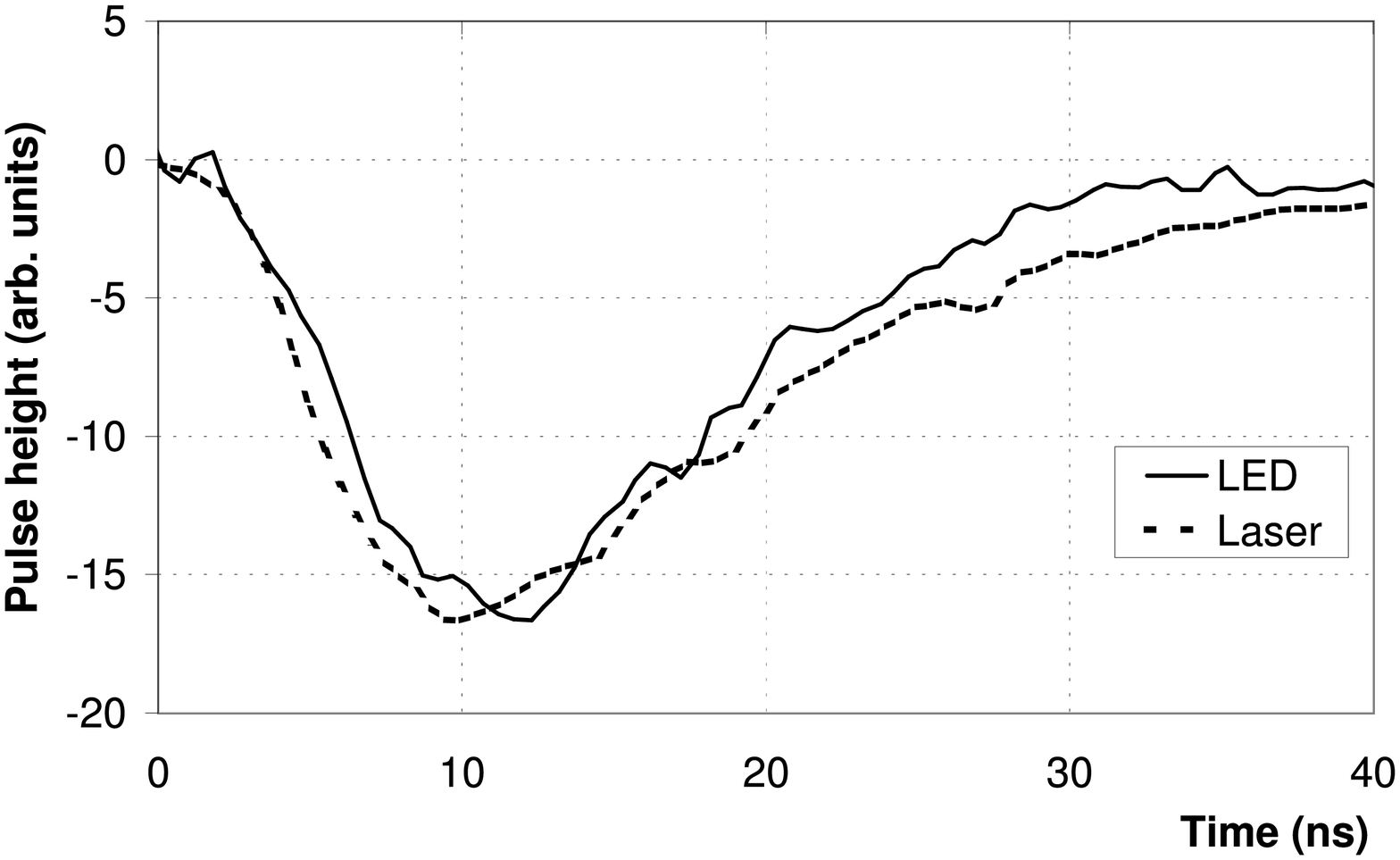}}
    \caption{Comparison of the average pulse shapes from WLS 
fibres excited by a blue LED and by laser-stimulated scintillator.}
    \label{fig:pulseshape}
  \end{center}
\end{figure}

The LEDs have a risetime of $\sim$ 3-10 ns, depending upon pulse 
height.  The fluorescence time of the WLS fibres is typically 
7-8 ns, and is specified to be less than 10 ns \cite{numi589}; 
the risetime of the signal pulses also increases by $\sim$ 0.7 
ns per metre of WLS fibre \cite{border01}.  
Fig.~\ref{fig:pulseshape} shows the time-averaged shapes of a 
series of optical pulses from the WLS fibres excited by a blue 
LED and by scintillator that has itself been stimulated by a 
nitrogen laser \cite{pulse_shape}.  In each case, the pulses 
liberated about four photoelectrons from the photocathode: this 
is approximately the signal size that we would expect from each 
end of a MINOS far-detector scintillator strip following the 
passage of a minimum-ionising particle through its centre.  
The pulse shapes are almost identical in the two cases, and they 
are dominated by the response of the WLS fibre.  In general, the 
risetime of the LED has little effect upon the overall pulse 
shape.

\subsection{Stability}

The short-term pulse-to-pulse variation of
an LED was measured 
by a PIN photodiode circuit with amplifier and a digital
oscilloscope. The standard deviation of the pulse height 
distribution, which is actually dominated by electronics noise 
from the measurement circuit, was found to be 1.1\% of the mean 
value. Therefore, when the response of the PMT to these optical 
pulses is measured, the spread in measured light levels is 
dominated entirely by photoelectron counting statistics; there 
is no need to account for LED variations on a pulse-by-pulse 
basis.

However, the LED light output does suffer occasional jumps 
(typically once every few weeks), and these can sometimes 
amount to changes of several percent.  An independent 
measurement of light intensity is therefore needed to monitor 
the output of the LED, and a PIN photodiode was chosen for this 
purpose.

\section{PIN photodiodes}
The PIN photodiode used to monitor the relative light output 
from the LEDs is the Hamamatsu S5971 \cite{hamamatsu}. 
This was chosen primarily because 
it offers a fast response while still
having a large enough surface area to accommodate the 1.0 mm 
diameter light-distribution fibres.  Some characteristics are 
listed in Table \ref{tab:pinspecs}.

\begin{table}[htb]
  \begin{center}
    \begin{tabular}{|l|l|}
\hline
Active area size & {1.2} {mm} diameter \\
Spectral response & {320} {nm} to {1020} {nm} \\
Dark current & {0.07} {nA} typ. \\
Temperature coefficient of dark current & 
$15\% /{}^{\circ}\mathrm{C}$\\
Photosensitivity @ {470} {nm} & {0.24} {A/W} \\
Temperature coefficient of photosensitivity & 
$<0.01\%/{}^\circ\mathrm{C}$ 
at laboratory temperatures\\
\hline
    \end{tabular}
    \caption{PIN photodiode specifications.}
    \label{tab:pinspecs}
  \end{center}
\end{table}

\subsection{Stability}

\begin{figure}[htb]
  \begin{center}
    \resizebox{0.8\textwidth}{!}{\includegraphics{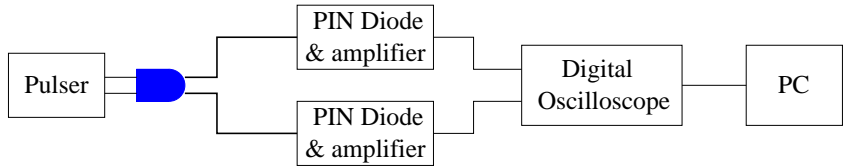}}
    \caption{Schematic diagram of the apparatus used to measure 
the PIN photodiode stability.}
    \label{fig:stabapp}
  \end{center}
\end{figure}

The light injection system is required to be stable as it 
tracks PMT gain changes over a period of about a month, during 
which time sufficient numbers of cosmic-ray muons will be 
detected to calibrate the detector.  The stability was tested 
by pulsing an LED continuously over a period of two months, 
and monitoring its output several times per second with a pair 
of PIN photodiodes.  After integration and shaping, these were 
read out  by
a digital oscilloscope (Fig.~\ref{fig:stabapp}).

\begin{figure}[htb]
  \begin{center}
    \resizebox{\textwidth}{!}{\includegraphics{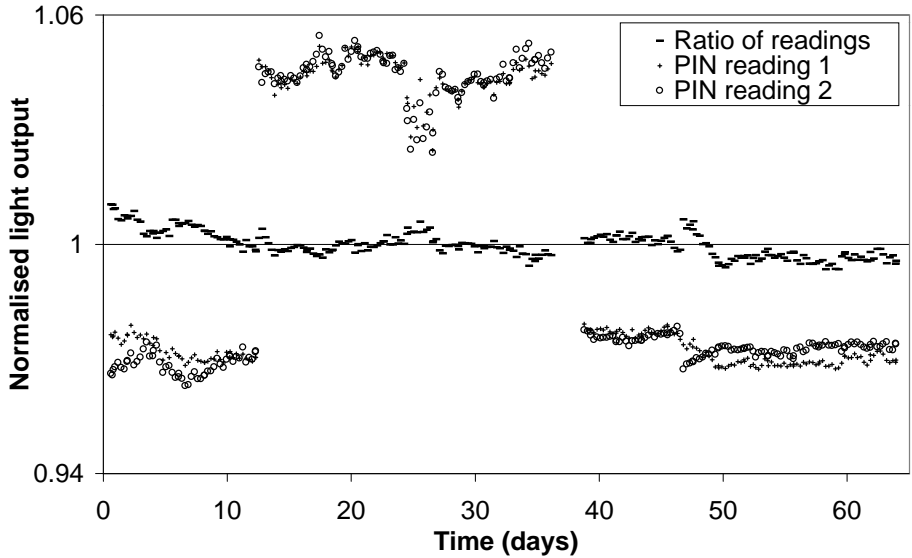}}
    \caption{Test of pin photodiode stability.  The PINs are stable
      to better than 0.5\% over two months.}
    \label{fig:pinstab}
  \end{center}
\end{figure}

The results of the stability test are shown in 
Fig.~\ref{fig:pinstab}. Each point in the figure is an average 
of a thousand consecutive 
readings: this is representative of the averaging that will 
take place during the MINOS calibration. For clarity, only 
one data point in sixteen is plotted. Over the first ten
days, the signal from PIN~1 gradually decreased; this is 
thought to be 
caused by the fibre settling slightly in its connector. 
The sudden jump in LED brightness at day 13 is unexplained, 
as is its instability in days 25-26; but the intensity 
changes were recorded by both PIN photodiodes, and the ratio 
of PIN outputs remained constant. A weekend power cut left 
the system off for an extended period in days 36-38; 
when restarted, the LED output returned to its original value.  
The feature at 46 days occurred when the
apparatus was accidently mechanically disturbed---a sharp 
deviation in one PIN's signal is again apparent, followed by a 
slow return to roughly the previous level.  

Although the light output of the LED changed by a few percent 
during the test period (e.g. the step changes at days 13 and 
37), the ratio between the individual PIN readings remained 
constant to within about 0.5\% over a period of two months: 
this is well within the MINOS stability specification.

\subsection{Linearity}

\begin{figure}[htb]
  \begin{center}
    \resizebox{0.9\textwidth}{!}{\includegraphics{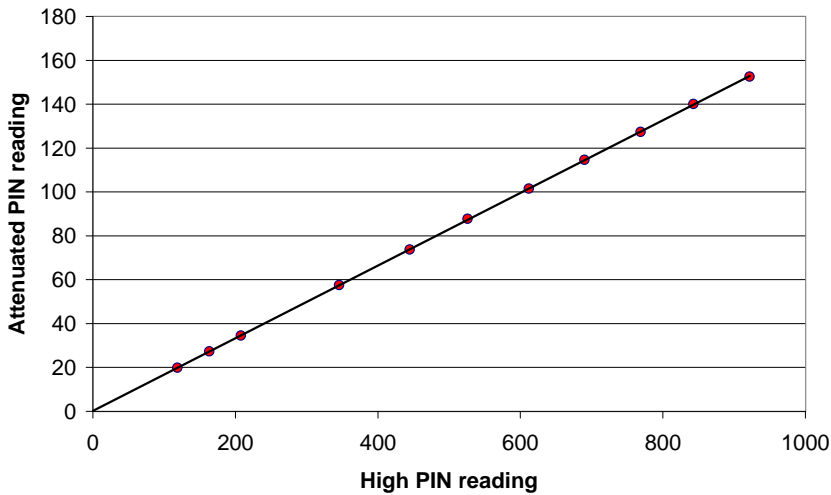}}
    \caption{PIN photodiode linearity. The plot shows the 
response in the PIN photodiode to light from the LED pulser 
passing through an optical attenuator, plotted against the 
signal with no attenuation, for several different pulse heights. 
The uncertainty on each 
point is dominated by the 8-bit readout of the digital 
oscilloscope, and is approximately 0.5\% of the full-scale 
reading on each axis.}
    \label{fig:pinlin}
  \end{center}
\end{figure}

The PIN photodiode must provide a scale that varies linearly with
applied light, in order to calibrate the non-linearities of the 
PMT.  The linearity of 
the PIN photodiode/amplifier system was measured by comparing 
the signal
from a pulsed LED with the signal from the same LED, but
with a neutral-density filter blocking most of the light. The 
amplified and shaped output from the PIN
photodiode was read out by a digital oscilloscope in peak-finding
mode. Repeating this measurement for several different pulse 
heights spanning the range of operation of the system 
produced the linear response shown in Fig.~\ref{fig:pinlin}.  
No evidence of non-linearity was observed.

The charge digitisation of the PIN photodiode pulse is also 
required to be linear. A charge-injection
system calibrates the ADC readout channels to an accuracy of 1\%,
thus ensuring that the PIN photodiode ADC scale is indeed 
proportional to the light input.

\subsection {Readout}

In the far detector, two PIN photodiodes are mounted on each of 
the front-end readout cards.  The current pulses produced by the 
PIN photodiodes are integrated and amplified, producing peak 
voltages that are a.c.\ coupled to the VA chips
to achieve the required current injection. The VA chip, whose 
primary purpose is to read out 
PMT pulses, requires a PMT dynode signal to trigger its readout. 
The optical wiring ensures that 
the PMT that triggers the readout of a given PIN photodiode will 
always be illuminated
by the same LED as the PIN photodiode in question. Each LED is 
in fact served by two PIN photodiodes, mounted on separate 
front-end readout cards; this
allows one readout card to be replaced (to change a PMT
or to replace an amplifier) whilst the LED calibration is held 
by the other PIN photodiode, which remains undisturbed. 

In the near detector, the phototubes are read out by QIE
electronics, which cannot be used to trigger the PIN readout. 
Instead, the readout of the PIN photodiodes is triggered by the 
presence of a signal on the dynode of a separate PMT 
(discussed in Section \ref{sec:trigPMT} below).

The optical fan-out from the LEDs is designed to deliver more 
light to the PIN photodiode than to the PMT pixels, in order to 
accommodate the relatively 
low sensitivity of the PIN.  In this way the full dynamic range 
of the signals
is well matched between PIN and PMT.  The signal-to-noise ratio of 
the PIN photodiode circuit is approximately 4000:1 at full scale.

\section{Trigger PMT}   
\label{sec:trigPMT}
At times, it will be necessary to take spectra of the PMT 
response at the single-photoelectron level.  In a significant 
fraction of events no signal will then be seen in the PMT, and 
consequently the PIN photodiode will not be read out.  In order 
to count these events properly, and in order to provide a 
distinctive identification to all light-injection events, each 
pulser box has an additional LED that emits a consistently 
bright flash with each pulse, regardless of the brightness of 
the primary LED pulse.  Light from this additional LED is carried 
to a dedicated M16 PMT -- the trigger PMT -- which is read out 
in the usual way, thus inserting a unique identifier in the 
data of each light-injection event.  In the near detector the 
trigger PMT serves an additional function, in that its dynode 
signal forces the readout of all PIN photodiodes.

\section{Mechanical design}
\subsection{Pulser boxes}
Each pulser box is 21 inches high, and fits in a standard 
19-inch rack. It is divided by a bulkhead wall into two 
compartments: one, at the front of the box, is for electronics; 
the other is for fibre optics.  The 20 LEDs are mounted on 
two circuit boards, supported on trays in the electronics 
compartment. A controller board,
containing the microprocessor and an RS232 interface, connects 
the two LED driver boards.  

\begin{figure}[ht]
  \begin{center}
    \resizebox{0.7\textwidth}{!}{\includegraphics{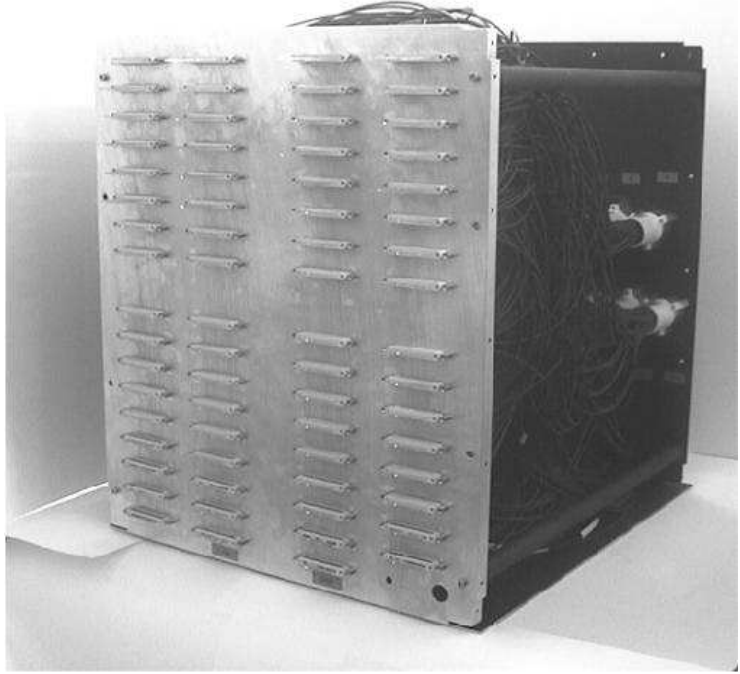}}
    \caption{The rear compartment of a pulser box, showing the 
back-panel  connectors. The side panels have been removed, and 
the collars at the end of each of the two rows are visible.  
A fibre from each collar runs to each 20-wide connector. 
Photograph courtesy OMC ltd.}
    \label{fig:pulser_rear}
  \end{center}
\end{figure}

The LEDs overhang the ends of their circuit boards and
pass through holes in the bulkhead into optical fan-outs 
known as cones. Each cone has a set of 71 optical fibres 
(including spares and dedicated, centrally-located fibres 
to carry light to PIN photodiodes).  These fibres take the 
optical signals from the cones to four columns of 16 
connectors on the back panel 
(Fig.~\ref{fig:pulser_rear}) 
of the box; each back-panel connector holds 20 fibres -- 
one from each cone.  There are three
additional 20-wide connectors on the back panel --- two 
provide light to the PIN photodiodes
(and so are fed by the dedicated PIN photodiode fibres), and 
the third is a spare. There is also a
single-fibre connector carrying light to the trigger PMT.  

\begin{figure}[ht]
  \begin{center}
    \resizebox{0.8\textwidth}{!}{\includegraphics{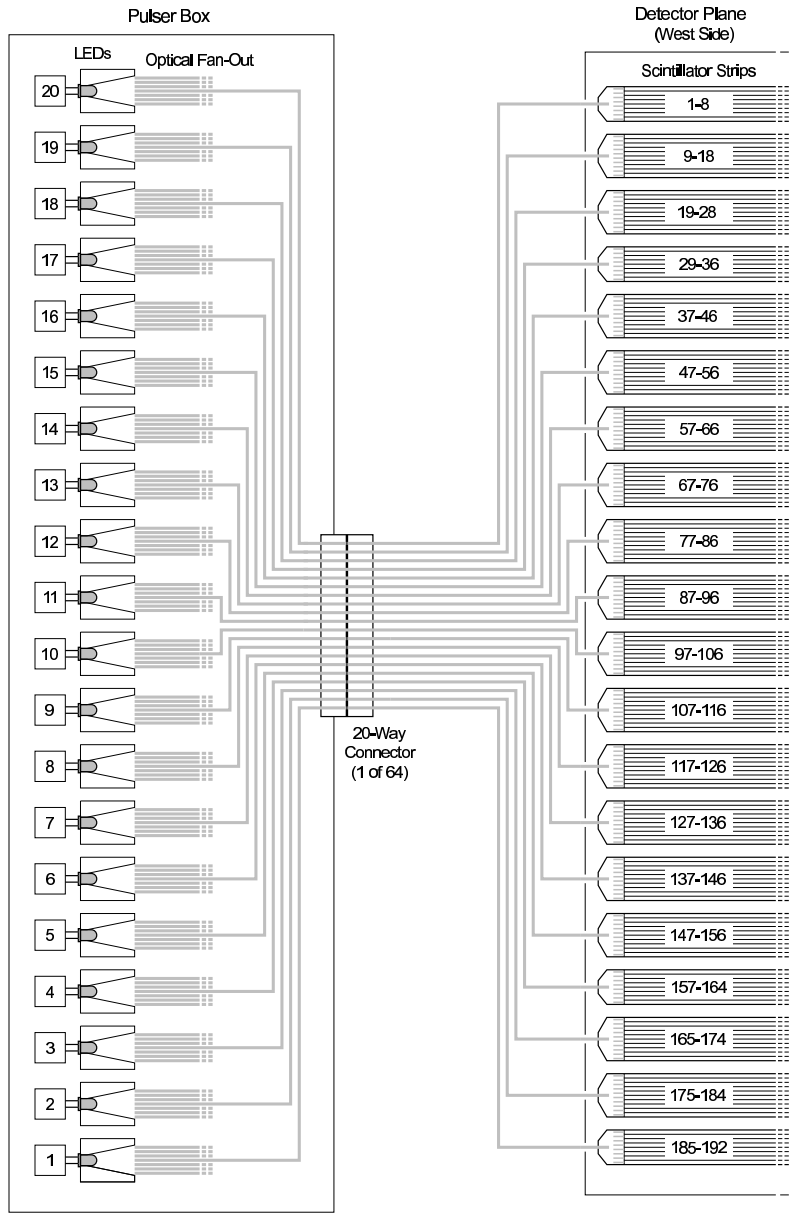}}
    \caption{Schematic showing typical optical routing for the 
far detector light injection system.  
On certain planes, multiplexing constraints require that the 
ordering of fibres within a bundle is changed: but the basic 
pattern wherein each of the 64 20-way connectors contains a 
fibre from each LED, and a fibre bundle from each such connector 
feeds in turn one half of a single detector plane, always holds.
}
    \label{fig:opt_routing}
  \end{center}
\end{figure}

Bundles of 20 fibres mounted externally to the box transport 
the light from each of the back-panel connectors to the 20 
light-injection points on each side of each far-detector plane; 
this is illustrated in Figure \ref{fig:opt_routing}.  The 
optical routing in the near detector is identical, except that 
there are fewer light-injection points on each plane (eleven 
for fully instrumented planes, seven for partially instrumented 
planes, and on one side of the detector only).  The calibration 
detector has just three light-injection points on each side of 
each of its 60 planes, and is thus served by six LEDs in a 
single pulser box.

\subsection{Cones and ponytails}

\begin{figure}[htb]
  \begin{center}
    \resizebox{\textwidth}{!}{\includegraphics{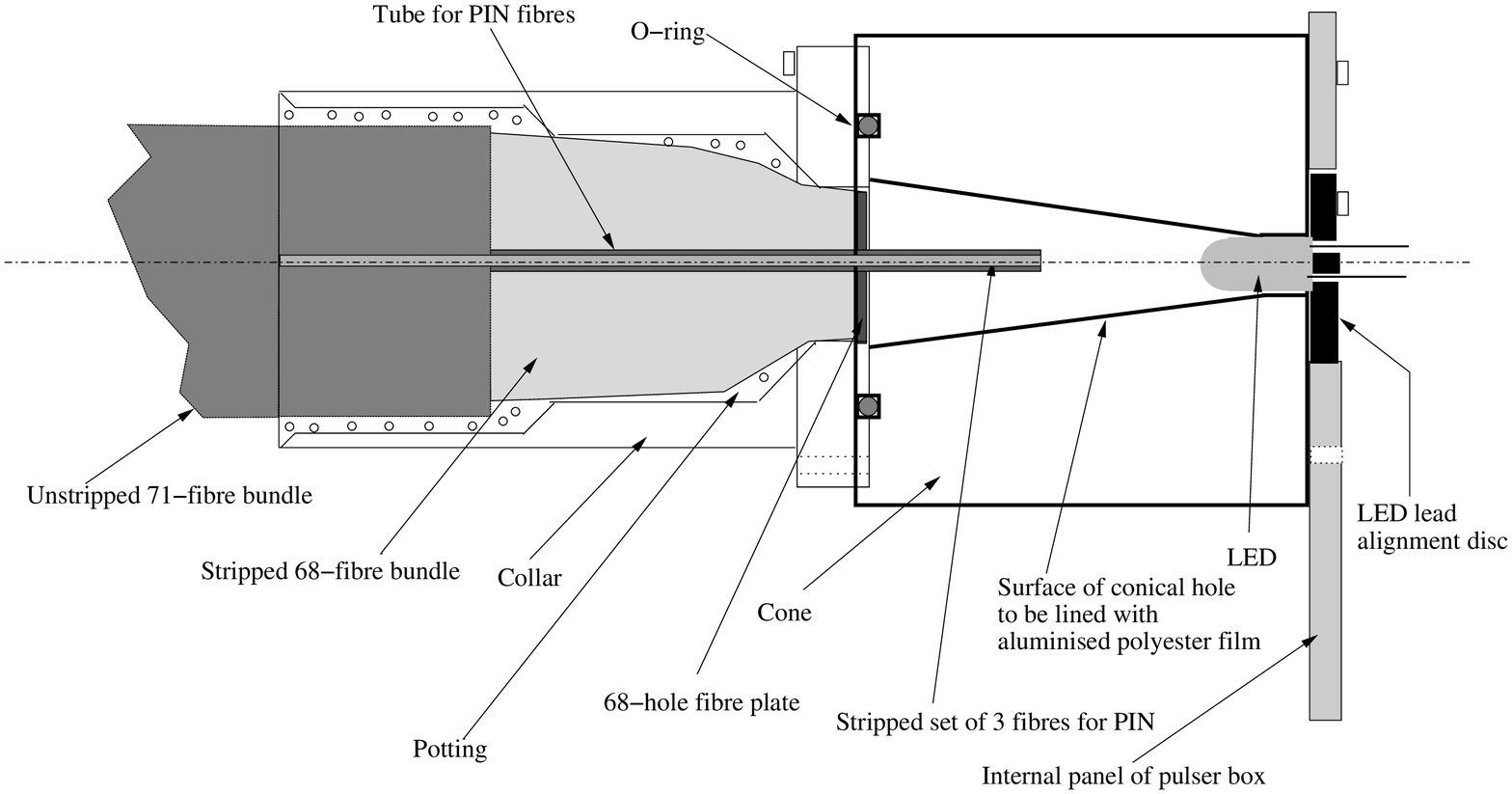}}
    \caption{Cone and collar assembly for pulser box.}
    \label{fig:coneass}
  \end{center}
\end{figure}

The basic design for the optical fan-out is a conical hole 
bored longitudinally through an aluminium cylinder, with the 
LED seated at the apex and the 71-fibre bundle attached at 
the base, as shown in Fig.~\ref{fig:coneass}. 
The walls of the cone are lined with aluminised
polyester (with a reflectivity greater than 99\%) to maximise 
the intensity of light delivered to the fibres and to assist 
with the uniformity of illumination of the fibre bundle.  
The fibre bundle, or ``ponytail'', is glued into a plastic 
collar which is attached to the cone. During assembly, the ends 
of the fibres are positioned precisely within the collar by 
means of a plastic disc with holes drilled in it, known as a 
``tea-strainer''. The surface of this is fly-cut with a 
diamond-tipped tool.

\begin{figure}[htb]
  \begin{center}
    \resizebox{0.7\textwidth}{!}{\includegraphics{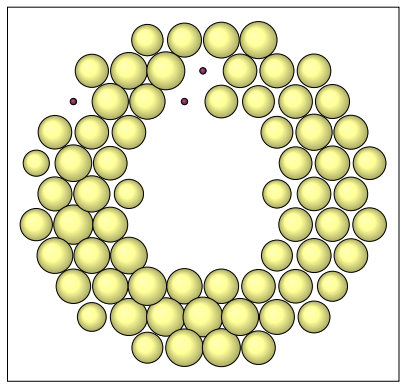}}
    \caption{Uniformity of illumination of a fibre tea-strainer 
with a wide-angle LED.  The area of each circle is proportional 
to the intensity of illumination.  The bundle included three 
spare fibres whose outputs were not measured.  }
    \label{fig:unif}
  \end{center}
\end{figure}

Three fibres (including one spare) are designated to serve the 
PIN photodiodes.  These are placed in the centre of the fibre 
bundle. To provide more light, these fibres are extended towards 
the LED, and are encased in a rigid plastic tube for support 
(see Fig.~\ref{fig:coneass}).  This also assists in reducing the 
non-uniformity of illumination of the light-injection fibres, as 
the PIN photodiode fibres occupy the area of brightest 
illumination.  Fig.~\ref{fig:unif} shows the uniformity of 
illumination of a typical fibre tea-strainer. In this example, 
the RMS deviation is approximately 15\% of the mean value, and 
the ratio of maximum to minimum intensities is 2.28. On later 
production models the uniformity improved somewhat, with the
ratio of maximum to minimum intensities often lying below 2.0.

  The PIN photodiode fibres are significantly longer than all 
other fibres in the ponytail (6 m vs.\ ~0.8 m).  This minimises 
the differences in relative arrival times of the optical pulses 
at the PIN photodiodes and at the PMT photocathodes, by 
compensating for the different optical path lengths: this ensures 
in turn that the amplified and shaped pulse from the PIN 
photodiode readout circuit is sampled and digitised at its peak.

\begin{figure}[htb]
  \begin{center}
    \resizebox{0.7\textwidth}{!}{
    \rotatebox{270}{\includegraphics{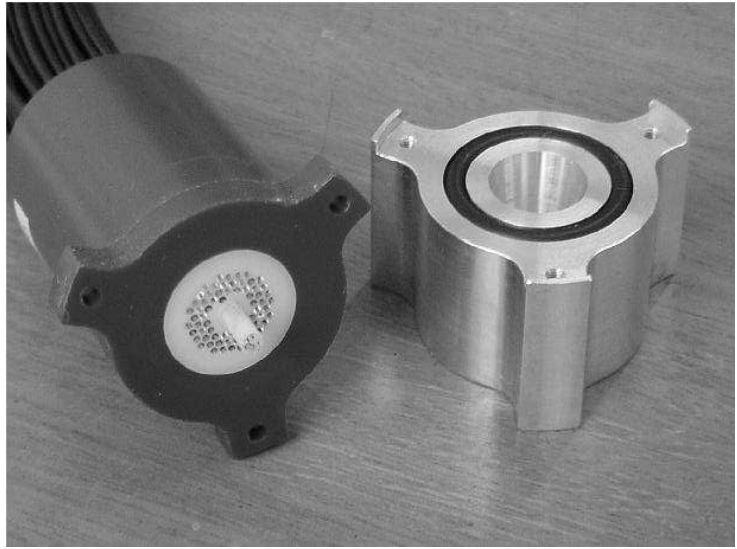}}}
    \caption{A collar and cone.  The tea-strainer and the 
central tube holding the PIN photodiode fibres are seen at the 
front of the collar, and the fibre bundle (ponytail) is emerging 
from its rear. }
    \label{fig:conecollar}
  \end{center}
\end{figure}

A disassembled cone and collar are shown in 
Fig.~\ref{fig:conecollar}.  

\subsection{Optical fibre}
The optical fibre used has a 1 mm diameter core of polymethyl 
methacrylate, with a fluorinated polymer cladding, and was 
manufactured by Mitsubishi Corp. \cite{mitsubishi}.  
The attenuation at 473 nm (the wavelength of the blue LEDs) 
is approximately 0.08 dB/m.  The fibre is jacketed with 2.2 mm 
diameter opaque black PVC that has a UL-VW1 \cite{ul} 
flame-resistance rating, a fire-safety requirement.  

\subsection{Optical connectors}
The multi-fibre connectors, both at the rear of the pulser 
box and on the mating ends of the external fibre bundles, are 
made of plastic.  Once the fibres have been glued into place, 
the connectors are fly-cut with a diamond-tipped tool: this 
provides extremely good optical transmission, of up to 90\% 
for each connector pair, without damaging the fibres and with 
a considerable saving in manpower over a hand-polishing process.  
The design for these connectors was adapted from that for the 
MINOS optical readout connectors.

Individual fibre ends in the light-injection system are all 
terminated with an H19 metal connector from OMC \cite{omc} 
(for those carrying light to the PIN photodiodes) or a 
custom-made, slightly larger variant thereof known as the 
CLRC01 (for those plugged into the light-injection points on 
the detector).  After gluing and light crimping into the 
connector, these fibre ends are hand-polished to optimise 
transmission.  

\section{Light injection modules}
\label{sec:lim}

\begin{figure}[ht]
  \begin{center}
    \resizebox{0.7\textwidth}{!}{\includegraphics{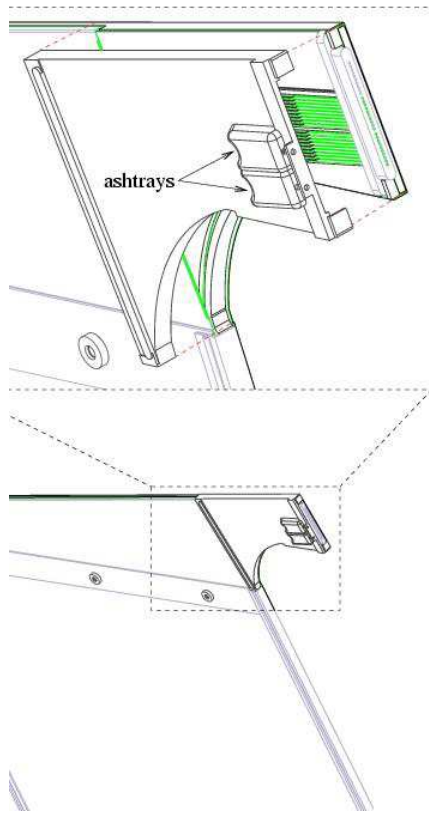}}
    \caption{Manifold parts. The light-injection module, here 
shown detached, forms the cover to the end of the manifold.  The 
exteriors of the two ``ashtray" compartments are marked.
    }       \label{fig:manifoldbits}
  \end{center}
\end{figure}

\begin{figure}[htb]
  \begin{center}
    \resizebox{0.7\textwidth}{!}{
    {\includegraphics{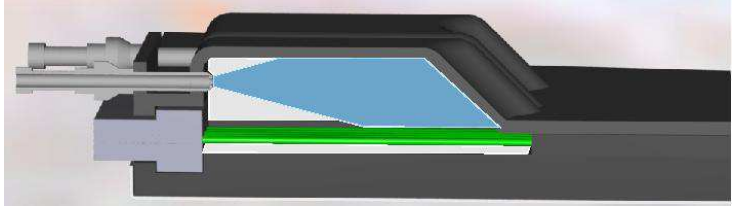}}}
    \caption{A cutaway computer model illustrating the concept 
behind the light-injection module. A curtain of ten green WLS 
fibres runs from right to left along the bottom of the LIM cavity, 
and the injected blue light illuminates them from above. The 
T-shaped component at the lower left is a cutaway of the bulk 
optical connector, into which the WLS fibres are glued.  In the 
upper left foreground is a cutaway of a CLRC01 connector, which 
terminates the light-injection fibre; an uncut connector is 
visible behind it.}
    \label{fig:limconcept}
  \end{center}
\end{figure}

\begin{figure}[hbt]
  \begin{center}
    \resizebox{0.7\textwidth}{!}{\rotatebox{270}{\includegraphics
    [clip=true, viewport = 0 0 400 630]{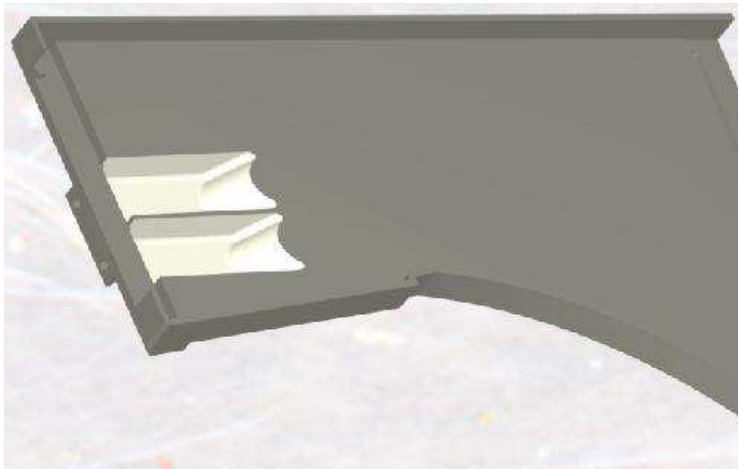}}}
    \caption{A 3D-rendered solid computer model of a 
light-injection module.  }
    \label{fig:lim}
  \end{center}
\end{figure}

At each end of the scintillator modules, there is a manifold
(Fig.~\ref{fig:manifoldbits}) that
draws the green WLS fibres from the scintillator strips together 
into a multi-fibre connector.  
This is the only point of access to the green fibres, and it is 
here that the LED light enters the optical readout chain.  The 
space is tightly constrained --- there is a plane of steel 
adjacent to the
scintillator module on one side, and another just 1 cm away on 
the other side. This does not allow room for the clear optical 
fibres carrying LED light to
arrive transversely to the planes: instead, they are brought in 
parallel to the readout fibres. The light illuminates a small 
($\sim$5 cm$^3$) compartment having diffusely-reflecting walls, 
across the top of which runs a ``curtain'' of eight or ten WLS 
fibres.  This is illustrated in Fig.~\ref{fig:limconcept}.  The 
WLS fibres absorb the blue LED light, which is then re-emitted 
isotropically in the green part of the spectrum.  Some of this 
green light is trapped in the fibre, and propagates in both 
directions towards the PMTs at either side of the far detector.  

The light-injection modules (LIMs; see Fig.~\ref{fig:lim}) 
containing the compartments (``ashtrays'') in which the 
absorption takes place are made from injection-moulded black 
plastic \cite{black_noryl}, and are mounted on the underside of 
each detector module, thus forming part of the manifold cover 
(see Fig.~\ref{fig:manifoldbits}). The 16- and
20-wide scintillator modules have two ashtrays in each LIM, and 
the wider modules have three. A complete MINOS far-detector plane 
contains 192 scintillator strips, whose green fibres are divided 
amongst a total of 20 ashtrays. Each ashtray is supplied with the 
blue light from a single plastic fibre, derived ultimately from 
one of the 20 LEDs in a pulser box.  Each of the LEDs illuminates 
the same ashtray in up to 64 planes.  

\begin{figure}[htb]
  \begin{center}
    \resizebox{0.9\textwidth}{!}{\includegraphics{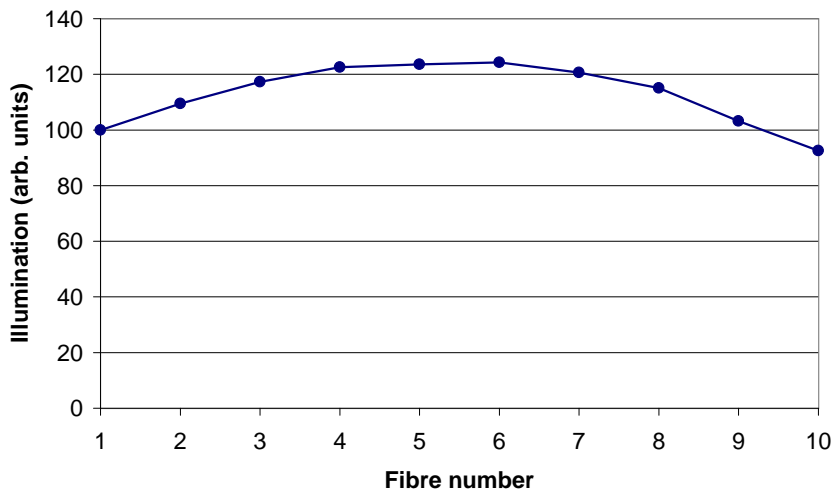}}
    \caption{Uniformity of illumination of the ten fibres in a 
light-injection module ``ashtray''.}
    \label{fig:unif_across_ashtray}
  \end{center}
\end{figure}

As discussed earlier, the total amount of light
illuminating each of the wavelength-shifting fibres should be as 
uniform as possible. The central position of the light-injection 
fibre in each ashtray results in a stronger illumination of the 
fibres at the centre than at the edges of the ashtray.  This 
tendency is countered to some extent by incorporating a curved 
reflecting back, as shown in Fig.~\ref{fig:lim}, which both 
shades a portion of the central fibres and reflects more light 
towards the edges than would a flat back. The uniformity of 
response across the ashtray is shown in 
Fig.~\ref{fig:unif_across_ashtray}. 
The walls of each ashtray are coated with a highly reflective 
$\mathrm{TiO}_2$-based paint (Bicron BC625).

\section{Optical routing constraints}

The far detector is eight-fold optically multiplexed. It is 
essential that only one fibre is
illuminated at any one time on any given pixel. As eight or ten 
adjacent fibres 
are illuminated simultaneously in an ashtray, this places a 
constraint on the 
multiplexing pattern used --- any scheme where adjacent strips are
mapped to the same pixel is not permissible.

\begin{figure}[htb]
  \begin{center}
    \resizebox{0.4\textwidth}{!}{\includegraphics[clip=true, 
viewport = 0 0 146 146]{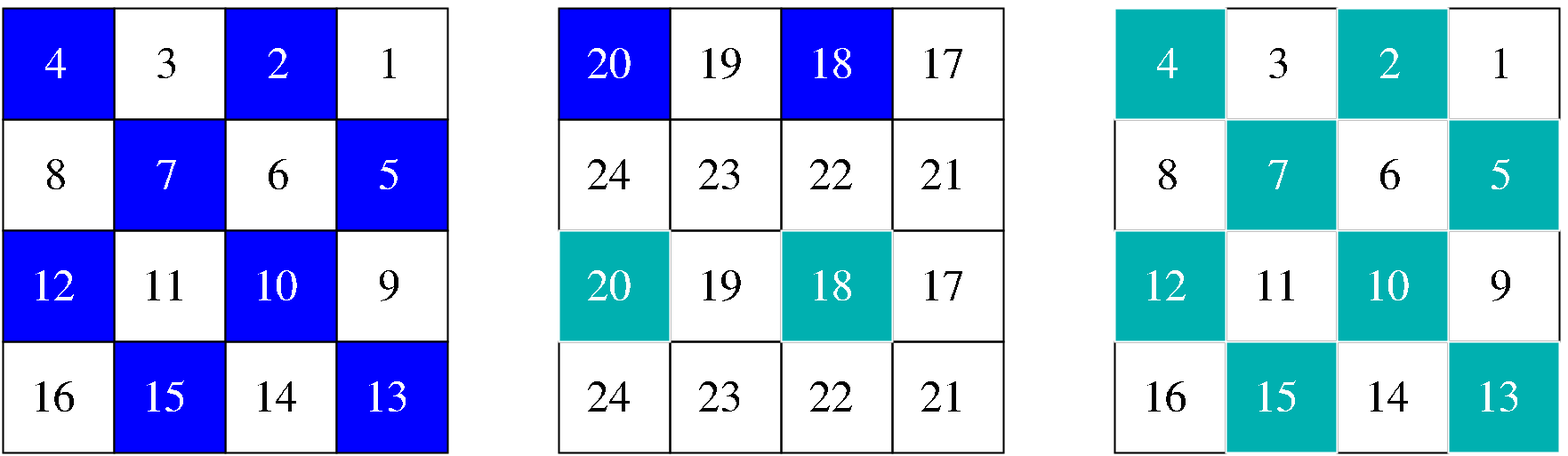}}
    \caption{Chessboard illumination pattern for M16s in the far
      detector. Pixels are numbered from right to left and 
top to bottom when viewing the face of the PMT. The shaded pixels 
are illuminated by a single ashtray, and are therefore illuminated 
simultaneously by light from the same LED.  Note that no two 
edge-adjacent pixels are illuminated at the same time. }
    \label{fig:m16checker}
  \end{center}
\end{figure}

It is also desirable to avoid simultaneous pulsing of adjacent 
PMT pixels, in order to mitigate the effects of crosstalk 
\cite{lang01}.  This is the reason for only illuminating up to 
10 WLS fibres (one ashtray) at any one time in any given plane.  
Arranging the fibres from sequential scintillator strips on 
alternate M16 pixels ensures that the pixels are illuminated in a
chessboard pattern, as in Fig.~\ref{fig:m16checker}, with diagonal 
rather than edge-adjacent pixels being illuminated simultaneously. 
The mapping is staggered on one side of the plane so that a hit 
can be located in the plane by using both ends of the readout.

These considerations lead to a simple algorithm for the optical 
wiring of the pulser boxes.  Each connector at the back of the 
pulser box contains one fibre from each of the 20 LEDs, and the 
20 fibres from each such connector feed the 20 ashtrays of one 
end of a single plane.  Thus, any given LED illuminates up to 
ten green fibres on each of 64 planes.

\section{Electronics}
The LED pulser offers the ability to change both the height 
and the width of the pulse electronically, by varying the 
current pulse that is applied to the LED. Each pulser box 
contains up to 20 LEDs, although in the near detector only 11 LEDs 
are used. As discussed in Section \ref{sec:trigPMT}, the 
controller card incorporates an
additional ``trigger'' LED that pulses at a fixed intensity and 
which is used to flag light-injection events in the data stream.

\subsection{Pulser control}

The pulser box is controlled by a PIC16F877 microprocessor 
\cite{microchip}. It communicates with a control PC via 
ASCII commands over a serial RS232
link. To avoid a proliferation of different types of data-carrying
cable running around the detector, the signals are carried over
the detector Ethernet, and converted for the pulser box by an
Ethernet-RS232 converter. In the standard mode of operation, the 
PC instructs the
system to pulse a given LED a certain number of times (typically 
1000) at a given height, width, and frequency (from 0.5 Hz to 
1 kHz). The microprocessor then controls the actual
pulsing. The pulser box is also supplied with an external trigger 
input, which forces a pulse of its current height/width setting 
in response to a TTL
signal. This can be used in the near detector to investigate the
effects of the timing of a pulse with respect to the 19 ns
time buckets in the QIE electronics.

\begin{figure}[htbp]
  \begin{center}
    \resizebox{0.5\textwidth}{!}{\includegraphics
    [clip=true, viewport = 136 278 422 736]
    {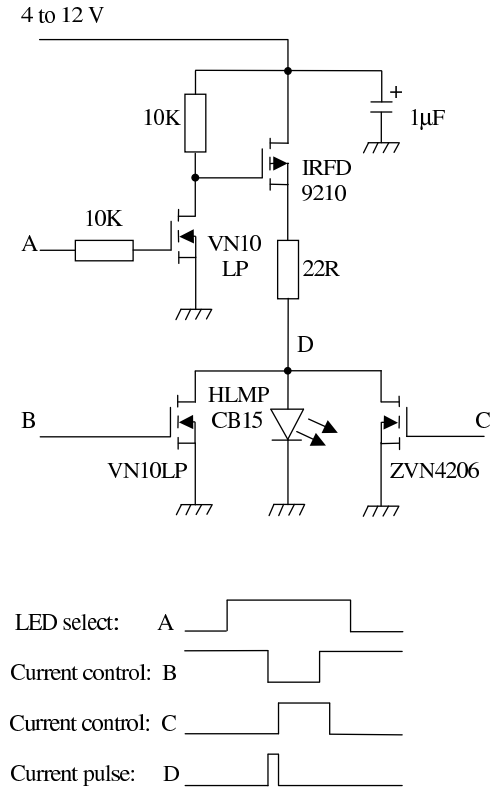}}
    \caption{LED pulser drive circuitry.}
    \label{fig:pulsercircuit}
  \end{center}
\end{figure}

Fig.~\ref{fig:pulsercircuit} shows the LED driver stage from the
pulser box circuit. The applied voltage, which determines the pulse height, is 
controlled by a 10-bit DAC. When the signal at A goes high, the
driver stage is turned on --- voltage is applied to the
22 $\Omega$ resistor, and the current is shunted to ground via
the left-hand FET.  This occurs typically a microsecond before the 
actual pulse, to give the current time to stabilise. (This 
``bypass'' current is turned off between each pulse in order to 
avoid excessive heating.) When the signal at B goes low, current 
flows through the LED, which emits light. A short time later, 
signal C goes high, and the current is shunted to ground via the 
right-hand FET; the LED is then turned off. The pulse width is 
determined by the delay between B turning off and C turning on; 
this is controlled by a programmable
delay line, giving a current pulse width of 15 to 36 ns in 3 ns
steps. The light pulses delivered to the phototube produce signals
ranging from a fraction of a photoelectron up to about
500 photoelectrons.

\section{Performance}

During the summer of 2001, the MINOS calibration detector was 
placed in a charged-particle test beam at CERN.  One of its 
functions was to confirm that the light-injection system could 
compensate for changing PMT gains at the 1\% level. The 
calibration detector was fully equipped with MINOS readout 
electronics.   

\begin{figure}[htbp]
  \begin{center}
    \resizebox{0.9\textwidth}{!}{\includegraphics{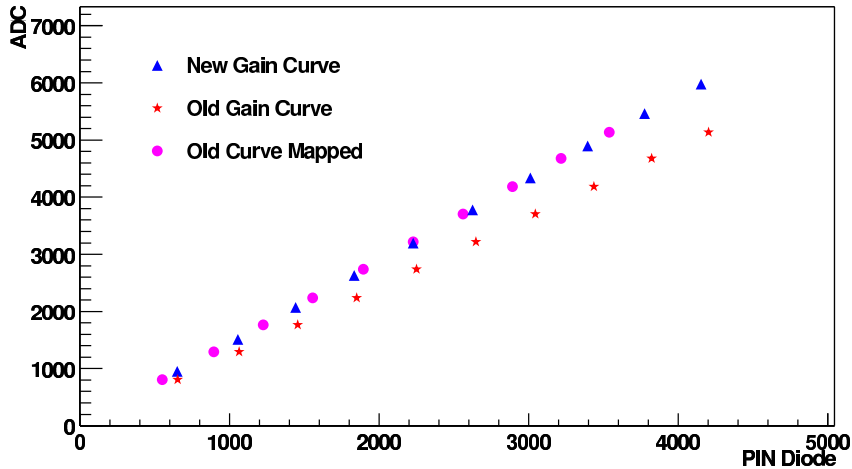}}
    \caption{Photomultiplier gain curves.  For several different 
intensities of injected light, the response of the PMT is shown 
as a function of the PIN photodiode output.  ``Old Gain Curve'' 
and ``New Gain Curve'' were taken at different values of the 
supply voltage; the light-injection system was used to provide 
a mapping from the old to the new curve, and thus to measure the 
large change in gain.}
        \label{fig:gain_curves}
  \end{center}
\end{figure}

The gain curves of the PMTs were produced by injecting light at 
ten different levels, and plotting the response of the PMTs as a 
function of the PIN photodiode output.  Figure 
\ref{fig:gain_curves} shows two such plots, labelled 
``Old Gain Curve'' and ``New Gain Curve'', measuring the same 
channel at different times. Each point represents the 
mean $\mu$ of 2000 optical pulses. The gain in general was 
observed to drift only slowly (less than 2\% over two months); 
but at one point the voltage supplied to the PMTs was increased, 
resulting in the dramatic difference between the two gain curves 
shown here.   

Drift in gain is monitored by looking at specific points on the 
gain curve on a regular (typically hourly) basis, as discussed in 
Section \ref{sec:aims} above.  These points are used to calculate 
a multiplicative factor for each channel that is applied to the 
entire  gain curve so as to mimic the time-evolution of the gain.  
The  underlying assumption is that the actual shape of the gain 
curve does not change on a short timescale. Figure 
\ref{fig:gain_curves} includes the recalculated gain-curve points.

\begin{figure}[htbp]
  \begin{center}
    \resizebox{0.9\textwidth}{!}{\includegraphics{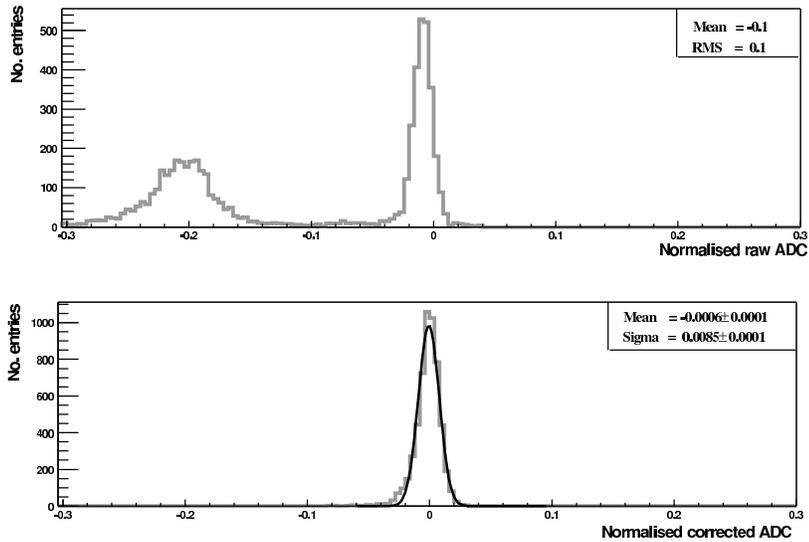}}
    \caption{Relative gain changes in the M16 photomultiplier 
tubes resulting from a change in the applied high voltage. The 
upper (lower) histogram shows the distribution before (after) 
calibration corrections are applied.  }
        \label{fig:calibworks2}
  \end{center}
\end{figure}

The performance of the light-injection system can therefore be 
tested by selecting different points on the gain curves, and 
comparing the mean ADC for a given light input with that 
predicted from the drift-point calculation.  The results of this 
procedure are shown in Figure \ref{fig:calibworks2}.  The 
upper histogram shows the relative difference between the means of 
measured points on separate gain curves, i.e. 
(ADC(0)-ADC($t$))/ADC(0), and the lower histogram shows the 
relative difference between the measured and corrected points, 
i.e. (ADC(0)$_{\rm corr}$-ADC($t$))/ADC(0)$_{\rm corr}$.  This 
technique is independent of any changes in light output from the 
LEDs; any change in gain of the PIN photodiodes would appear as 
an offset from zero of the lower histogram.  The width of the 
lower histogram is limited by photoelectron counting statistics 
at each data point: the ultimate precision achievable by the 
system is not yet completely determined, but the figure 
demonstrates that the calibration procedure has corrected the 
gain changes to within the 1\% specification required by MINOS.  

\section{Conclusion}

A light-injection system has been developed to assist in the 
calibration of the MINOS detectors.  Based upon blue LEDs, 
with PIN photodiodes to monitor relative intensities, the system 
is able to measure non-linearities in PMT response as well as 
measuring changes in gain at the 1\% level on an hourly basis 
over timescales of a month or more.  

\section*{Acknowledgements}

This work has been funded by the UK Particle Physics and 
Astronomy Research Council and the US Department of Energy.  
The authors would like to thank the other members of the 
MINOS collaboration for their support, and in particular 
Jim Musser and Karol Lang, both of whom have provided valuable 
criticism of the document, and the members of the scintillator, 
DAQ, electronics, and integration/installation groups for all of 
the time and effort spent in accommodating the needs of the 
light injection system. Jim Grozier gave generously of his own 
free time to carry out crucial laboratory tests at Sussex. 
The assistance of CERN and its PS staff with the test-beam 
experiments is gratefully acknowledged.  The staff at OMC 
have also been most helpful.  This work is based in 
part on the dissertation of P.\ Adamson, submitted to the 
University of Sussex in partial fulfilment of the requirements 
for the D.Phil.\ degree \cite{adamson_phd}.

\bibliography{minos}

\end{document}